\documentclass[a4paper,fleqn,12pt]{article}
\usepackage{amssymb,amstext,latexsym,xspace,ifthen}
\usepackage{epsfig,color}

\newboolean{draft}          
\setboolean{draft}{false}
\newboolean{title}          
\setboolean{title}{true}
\newboolean{ack}            
\setboolean{ack}{true}
\newboolean{internal}            
\setboolean{internal}{false}

\ifthenelse{\boolean{draft}}{
\usepackage{drafthead}
\usepackage{showkeys}
}{} 

\makeatletter
\newif\if@fewtab\@fewtabtrue
\makeatother

\newcommand{\Eq}[1]{ Eq.~(\ref{#1})}
\renewcommand{\[}{\begin{eqnarray}}
\renewcommand{\]}{\end{eqnarray}}

\newcommand{\non}{\nonumber \\ }
\newcommand{\ft}[2]{{\textstyle {\frac{#1}{#2}} }}

\newcommand{\On}{\ensuremath{\mathbb{O}}\xspace}
\newcommand{\Rn}{\ensuremath{\mathbb{R}}\xspace}
\newcommand{\Zn}{\ensuremath{\mathbb{Z}}\xspace}

\newcommand{\cI}{\mathcal{I}}
\newcommand{\cL}{\mathcal{L}}
\newcommand{\cM}{\ensuremath{\mathcal{M}}\xspace}
\newcommand{\cV}{\mathcal{V}}

\newcommand{\SLR}[1][2]{\ensuremath{SL(#1,\Rn)}\xspace}
\newcommand{\SLZ}[1][2]{\ensuremath{SL(#1,\Zn)}\xspace}

\newcommand{\USp}[1][8]{\ensuremath{USp(#1)}\xspace}

\newcommand{\4}[1][{(4)}]{\ensuremath{F_{4 #1}}\xspace}
\newcommand{\6}[1][{(6)}]{\ensuremath{E_{6 #1}}\xspace}
\newcommand{\7}[1][{(7)}]{\ensuremath{E_{7 #1}}\xspace}
\newcommand{\8}[1][{(8)}]{\ensuremath{E_{8 #1}}\xspace}
\newcommand{\rep}[1]{\ensuremath{\mbox{\mathversion{bold}$\mathbf{#1}$%
                     \mathversion{normal}}}}

\def\moth{\mathsurround=0pt}
\newdimen\zo \zo=0pt

\def\tick{\leaders\hrule height 0.5ex depth 0pt \hskip 0.5pt}
\def\upboxfill{$\moth \setbox\zo\hbox{\tick}%
  \hskip 2pt\hbox to 0pt{$\tick$\hss}\hrulefill \hbox to 2pt{$\tick$\hss}$}

\def\dtick{\leaders\hrule height .34pt depth 0.5ex \hskip 0.5pt}
\def\downboxfill{$\moth \setbox\zo\hbox{\dtick}%
  \hskip 2pt\hbox to 0pt{$\dtick$\hss}\hrulefill%
  \hbox to 2pt{$\dtick$\hss}$}
\def\overbox#1{\mathop{\vbox{\moth\ialign{##\crcr\noalign{}
\downboxfill\crcr\noalign{\vskip 1pt\nointerlineskip}
      $\hfil\displaystyle{#1}\hfil$\crcr}}}\limits}

\newcommand{\oversym}[1]{\!\overbox{{}#1}}

\begin{document}

\ifthenelse{\boolean{title}}{
\thispagestyle{empty}
\renewcommand{\thefootnote}{}

\begin{flushright}
hep-th/0205235\\
KCL-MTH-02-13
\end{flushright}
\bigskip

\begin{center}
\mathversion{bold}
{\bf\Large Lifting U-Dualities}
\bigskip\bigskip\bigskip
\mathversion{normal}

{\bf\large K.~Koepsell and F.~Roose\medskip\\ }
{\em Department of Mathematics\\
King's College London\\
Strand, London WC2R 2LS, U.K.}
\footnote{email: {\tt koepsell@mth.kcl.ac.uk, fredr@mth.kcl.ac.uk}}

\end{center}
\bigskip
\medskip

\begin{abstract}
  We present a novel global \7 symmetry in five-dimensional maximal
  supergravity as well as an \8 symmetry in $d\!=\!4$. These symmetry groups
  which are known to be present after reduction to $d\!=\!4$ and $d\!=\!3$,
  respectively, appear as conformal extensions of the respective well-known
  hidden-symmetry groups. A global scaling symmetry of the Lagrangian is the
  key to enhancement of \6 to \7 in $d\!=\!5$ and \7 to \8 in $d\!=\!4$.  The
  group action on the physical fields is induced by conformal transformations
  in auxiliary spaces $\cM$ of dimensions $27$ and $56$, respectively. The
  construction is analogous to the one where 
  the conformal group of Minkowski space acts on the boundary of 
  AdS$_5$ space. A geometrical
  picture underlying the action of these ``conformal duality groups'' is
  given.
\end{abstract}

\vfill

\leftline{{\sc May 2002}}
\newpage
\renewcommand{\thefootnote}{\arabic{footnote}}
}{} 

\section{Introduction}\label{sec:introduction}
In the late 70s, extended supergravity theories in $d$ dimensions, $(d \leq
9)$ were demonstrated to exhibit certain continuous global symmetries, termed
`hidden symmetries' at the time~\cite{CreJul79}. More specifically, they were
found as invariances of the equations of motion supplemented with the Bianchi
identities. It was not until '94 that Hull and Townsend conjectured a discrete
version of these symmetry groups to persist as exact symmetries of
(compactified) type II string theories~\cite{HulTow95}. These discrete groups 
have become known since as `U-duality groups'.

Compactification on a $d$-torus relates the maximally-extended
supergravity in $11\!-\!d$ dimensions to the unique eleven-dimensional
one~\cite{CrJuSc78}.  The hidden-symmetry group in $1\le 11\!-\!d\le9$
dimensions is observed to be $E_{d(d)}$\footnote{For $d<6$ this is a classical
  group (see e.g. \cite{CJLP98}).}.
  
Despite this observation, it is a non-trivial question whether global
symmetries that become manifest only in lower-dimensional theories might be
secretly present nonetheless in higher dimensions. Indeed, in
Refs.~\cite{WitNic86,Nico87}, eleven-dimensional supergravity was reformulated
such that the space-time tangent-space $SO(1,10)$ symmetry was being replaced
by the local symmetries that otherwise would become manifest only upon
reduction to four and three dimensions, respectively. In these constructions
the dependence on all 11 space-time coordinates is retained; hence, the
symmetries are already present {\em before} dimensional reduction. Thus, part
of the hidden symmetries of dimensionally reduced supergravity can be
``lifted'' to eleven dimensions. Moreover, some evidence has been given that
the full \8 duality group~\cite{MelNic98,KoNiSa00} and even the Kac-Moody
algebra $E_{11}$~\cite{West01} can be realised in the 11-dimensional theory
hinting at the existence of an ``exceptional geometry''~\cite{MelNic98} in
eleven dimensions.

In the present paper, we describe a procedure to lift manifest global
symmetries up by one dimension. Rather than restricting ourselves to the 
maximal compact subgroup, we do not meet any obstacles to perform the lift 
for the {\em entire} group $G$. The key object in the construction is an 
auxiliary space, $\cM$,
on which $G$ has a natural nonlinear action. This nonlinear action is
most easily derived by viewing $\cM$ as the `boundary' of a submanifold that
is appropriately embedded in a larger space $\cM^\#$, on which $G$ acts {\em
  linearly}. As will be clarified, the steps involved are, in spirit,
reminiscent of those leading to the standard nonlinear action of the conformal
group in Minkowski space, whereby the latter is identified with the boundary
of AdS.

Further, the idea of this additional (besides the space-time dimensions, that
is) space was already present in Ref.~\cite{WitNic00}. Here, however, the new
ingredient is that the supergravity fields will now be allowed to live on 
$\cM$, even though they are propagating only in space-time. Therefore, these 
extra dimensions are not to be considered truly physical. Rather, the underlying
geometry of ${\cal M}$ provides a means to make an otherwise hidden symmetry manifest, via
the induced action on the fields in the theory.

The idea of making seemingly nontrivial dualities manifest via some
underlying geometry can hardly be called new: F-theory~\cite{Vafa96}, where
the \SLZ symmetry of a class of compactified IIB models is geometrically
understood as the modular group acting on an auxiliary two-torus (the elliptic
fibre), provides a prototypical example. Note that also there, no physical
significance is assigned to the added two torus directions.

The structure of the paper is as follows: in section~\ref{sec:confrel} we
outline the concept of conformal realisations. Starting with the
familiar example of the conformal group in Minkowski space in
section~\ref{sec:mink}, we apply the same construction to \7 acting 
conformally on a 27-dimensional space $\cM$ in section~\ref{sec:confext}; 
next, in section~\ref{sec:geometry} we put the
relation between the conformal \7 duality in $d\!=\!5$ and the linearly
realised \7 in four dimensions in a geometrical perspective. 
In section~\ref{sec:sugras}, we exploit the construction to find a realisation
of \7 in five-, and \8 in four-dimensional maximal supergravity theories. 
First, we review the known structure of five-dimensional supergravity in
section~\ref{sec:d5sugra}, and point out the minimal consistency conditions
that a true ``lifted'' symmetry must satisfy. Next, 
section~\ref{sec:e7realisation} introduces the 27-dimensional auxiliary
space $\cM$. An analysis of the consistency requirements fixes a peculiar
dependence of all fields on $\cM$ only via the cubic \6-norm $N$. Moreover, an
\7 embedding in the diffeomorphism group of $\cM$, namely, as the conformal
group w.r.t. $N$, is demonstrated to be consistent. Some additional
remarks are collected in section~\ref{sec:remarks}; section~\ref{sec:d4sugra},
finally, contains a similar construction resulting in an \8 conformal duality
symmetry in $d\!=\!4$.

\section{Conformal realisations}\label{sec:confrel}

As will become clear shortly, the nonlinear realisation of \7 in a
27-dimensional flat space bears striking similarities to that of the conformal
group $Conf(M^{3,1})$ in four-dimensional Minkowski space, $M^{3,1}$. Since
the latter has a clearer geometrical picture, it will serve as a model
throughout the paper.

\subsection{The conformal group in $M^{3,1}$}\label{sec:mink}

Minkowski four-space, $M \equiv M^{3,1}$, is the (flat) vector space
$\mathbb{R}^4$ endowed with the indefinite metric $\eta$. On $M$, the
conformal group $Conf(M^{3,1})$ consists of elements $g$ with an action on $x
\in M$ that preserves the lengths of vectors up to a scale factor:
\begin{equation}
\eta\,(g x, g x) = \lambda_g(x)\, \eta(x, x)\ .
\end{equation}
As is well-known, this condition singles out the transformations whose
infinitesimal form is given by:
\begin{eqnarray}
\delta_e x^\mu       &=& e^\mu \,,\label{mink-transl}\\[.5ex]
\delta_\Lambda x^\mu &=& \Lambda^\mu{}_\nu x^\mu + h\,x^\mu\,, 
\label{mink-lorentz}\\[.5ex]
\delta_f x^\mu       &=& 2\,(x^\nu f_\nu)\,x^\mu - (x^\nu x_\nu)\,f^\mu\,,
\label{mink-conf}
\end{eqnarray}
A word about notation: $e^\mu, f^\mu \in M^{3,1}$ parametrise translations and
so-called special conformal transformations\footnote{The integrated version of
  these are best thought of as a succession of inversion $(x^\mu \rightarrow
  \frac{x^\mu}{\eta(x,x)})$, translation, and inversion again.}, respectively,
while \Eq{mink-lorentz} contains the Lorentz transformations and dilatations,
parametrised by (antisymmetric) $\Lambda$ and $h\in\Rn$.  The given
transformations enlarge the Poincar\' e algebra to the conformal algebra
$so(4,2)$, which has a three-grading
\[
\begin{array}{ccc@{\,\,\oplus\,\,}c@{\,\,\oplus\,\,}c}
so(4,2) &=& \mathfrak{g}^{-1} & \mathfrak{g}^{0} & \mathfrak{g}^{+1}\\[1ex]
        & & (e^\mu)             & (\Lambda^\mu{}_\nu,h) & (f^\mu)
\end{array}
\label{3-grading}
\]
i.e., the degree of the commutator of two elements equals the sum of their
degrees. For example, Lorentz rotation generators have degree 0. Note further
that the infinitesimal transformations in grade $0$ act linearly, while grade
$+1$ elements have a quadratic action on the coordinates.

\mathversion{bold}
\subsection{Conformal extension of \6 to \7}\label{sec:confext}
\mathversion{normal}

In order to extend the global symmetry group \6 in five dimensions to \7 we
will introduce a 27-dimensional auxiliary space $\cM$ which admits \6 as
``generalised Lorentz group''. Like in Minkowski space, we
will extend this group to a ``generalised conformal group''. In fact,
to introduce such generalised space-times and the action of the corresponding
conformal group is an old idea~\cite{Guna75,Guna80}. More recently, it was
already suggested to extend space-time by extra dimensions~\cite{WitNic00}.

We shall make use of the following branching rule for the adjoint
representation of \7 with respect to \6
\[
\begin{array}{ccc@{\,\,\oplus\,\,}c@{\,\,\oplus\,\,}c}
\rep{133} &\stackrel{\6}\longrightarrow& \rep{27} & 
           \left[ \rep{78} \oplus \rep{1} \right]& 
           \rep{\overline{27}}\\[1ex]
&&\mathfrak{g}^{-1} & \mathfrak{g}^{0} & \mathfrak{g}^{+1}
\end{array}\ .
\label{e7-grading}
\]
The \6 adjoint \rep{78} acts linearly
on $\cM$, the carrier space of the
fundamental \rep{27} representation; $\cM$ can be endowed with a triple norm
$N_3$ that is preserved under \6, analogously to $\eta$ being Lorentz
invariant in Minkowski space.  Moreover, \Eq{e7-grading} displays the
three-graded structure of \7, similar to that of $Conf(M^{3,1})$ in Minkowski
space (see\Eq{3-grading}), and thus hints towards viewing \7 as a conformal
extension of \6.  Indeed, following the construction of Ref.~\cite{GuKoNi00},
one may define the following operations on $\cM$ with coordinates $Y^m$:
\begin{eqnarray}
\delta_{27}\, Y^m &=& E^m\,,\label{e7-trans}\non[.5ex]
\delta_{78}\, Y^m &=& {\Lambda^m}_n Y^n\,,\label{e7-rot}\non[.5ex]
\delta_{1}\,  Y^m &=& H\,Y^m\,,\label{e7-dil}\non[.5ex]
\delta_{\overline{27}}\, Y^m &=& \ft12 F_n K^{mn}{}_{qr} Y^qY^r \,,
\label{e7-conf}
\end{eqnarray}
These transformations define a nonlinear realisation of \7 on $\cM$,
provided that the coefficients $K^{mn}{}_{pq}$ are identified with the
structure constants of the \6-invariant triple product in the \rep{27}
representation of \6 (see Appendix~\ref{app:e6})\footnote{The explicit form of
  the coefficients $K^{mn}{}_{qr}$ can be found in Ref.~\cite{GuKoNi00}.}.
These are a generalisation of the transformations
in\Eq{mink-transl}~-\Eq{mink-conf}.  Moreover, the first two types of
transformations leave $N_3(Y - Y')$ invariant, while this quantity gets
rescaled only under transformations of the latter two types. As such, \6 acts
on $\cM$ as generalised Lorentz rotations, and the $-1$ ($+1$) subspace as
translations (special conformal transformations). It is thus fair to say that
\Eq{e7-trans}-\Eq{e7-conf} define a {\em conformal realisation} of \7 on a
27-dimensional space.

\subsection{Hidden exceptional geometry}\label{sec:geometry}

In $d\!=\!4, N\!=\!8$ supergravity, the gauge vectors and their
electric-magnetic duals combine into the $\mathbf{56}$ linear \7
representation. To see the relation with the $d\!=\!5$ situation, where \7
will be realised on $\mathbf{27}$, the geometric perspective adopted in this
section may give additional insight.

For simplicity, we start by reviewing the extension of the Lorentz $so(3,1)$
to the conformal $so(4,2)$ algebra in Minkowski four-space.  First, $so(4,2)$
is linearly realised on a six-dimensional vectorspace $M^{4,2}$, endowed with
a metric $\hat \eta$ of signature $(4,2)$.  In coordinates $X$, the following
equation defines a codimension-1 subspace:
\begin{equation}\label{eq:ads}
\hat\eta\, (X,X) = R^2\ ,
\end{equation}
for arbitrary fixed $R \in \mathbb{R}$. The thus-defined five-manifold with
the induced metric $g$ is identified as AdS$_5$. Since the linear action on
$M^{4,2}$ leaves the defining equation\Eq{eq:ads} invariant, so$(4,2)$
descends to the isometry group of $g$.  Alternatively, the fact that AdS$_5
\simeq SO(4,2)/ISO(3,1)$ makes this property manifest.

\begin{figure}\label{fig:AdS}\begin{center}
\leavevmode
%
%
\begin{picture}(0,0)%
\includegraphics{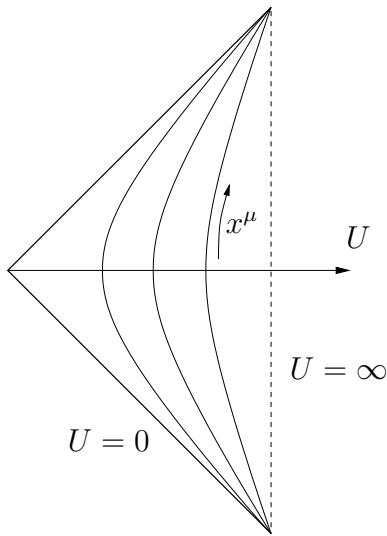}%
\end{picture}%
\setlength{\unitlength}{2072sp}%
\begingroup\makeatletter\ifx\SetFigFont\undefined
\def\x#1#2#3#4#5#6#7\relax{\def\x{#1#2#3#4#5#6}}%
\expandafter\x\fmtname xxxxxx\relax \def\y{splain}%
\ifx\x\y   
\gdef\SetFigFont#1#2#3{%
  \ifnum #1<17\tiny\else \ifnum #1<20\small\else
  \ifnum #1<24\normalsize\else \ifnum #1<29\large\else
  \ifnum #1<34\Large\else \ifnum #1<41\LARGE\else
     \huge\fi\fi\fi\fi\fi\fi
  \csname #3\endcsname}%
\else
\gdef\SetFigFont#1#2#3{\begingroup
  \count@#1\relax \ifnum 25<\count@\count@25\fi
  \def\x{\endgroup\@setsize\SetFigFont{#2pt}}%
  \expandafter\x
    \csname \romannumeral\the\count@ pt\expandafter\endcsname
    \csname @\romannumeral\the\count@ pt\endcsname
  \csname #3\endcsname}%
\fi
\fi\endgroup
\begin{picture}(4129,6344)(1329,-6833)
\put(3961,-3211){\makebox(0,0)[lb]{\smash{\SetFigFont{12}{14.4}{rm}{\color[rgb]{0,0,0}$x^\mu$}%
}}}
\put(2071,-5821){\makebox(0,0)[lb]{\smash{\SetFigFont{12}{14.4}{rm}{\color[rgb]{0,0,0}$U=0$}%
}}}
\put(5401,-3391){\makebox(0,0)[lb]{\smash{\SetFigFont{12}{14.4}{rm}{\color[rgb]{0,0,0}$U$}%
}}}
\put(4726,-4966){\makebox(0,0)[lb]{\smash{\SetFigFont{12}{14.4}{rm}{\color[rgb]{0,0,0}$U=\infty$}%
}}}
\end{picture}
%
%
\caption{AdS space in Poincar{\'e} coordinates}
\end{center} \end{figure}

A coordinate system on AdS$_5$ that will prove particularly useful, is that of
the so-called Poincar{\' e} coordinates $(x^\mu, u), \mu = 0\ldots 3$. They
enjoy the following properties:
\begin{enumerate}
\item the subalgebra $iso(3,1)$ of isometries has a linear realisation on
  $(x^\mu)$;
\item the dilatation descends to a rescaling $(\lambda x^\mu, \lambda^{-1} u)$.
\end{enumerate}
In these coordinates, AdS$_5$ is a foliation parametrised by $u$.  Moreover,
Minkowski space arises in this picture as the boundary of AdS$_5$: $\{u =
\infty\}$, and special conformal transformations are nonlinearly realised on
$(x^\mu)$ (as given in \Eq{mink-conf}). In summary, the picture outlined
picture yields an understanding how transformations that initially act
linearly get translated into nonlinear ones, now realised on the boundary of
an invariantly embedded submanifold, though.

Next, when \7 is viewed as a conformal extension of \6, a closely parallel
geometric reasoning will lead to the desired relation between the
$\mathbf{56}$ and $\mathbf{27}$ realisations. The branching rule
\begin{eqnarray}
\mathbf{56} 
&\stackrel{\6}\longrightarrow& \mathbf{1} \oplus \mathbf{27} \oplus 
  \overline\mathbf{27} \oplus \overline{\mathbf{1}}\ .
\end{eqnarray}
suggests a choice of coordinates on the carrier space $\cM_{56}^\#$ of
$\mathbf{56}$ that reflect this decomposition: $\hat{Z} := (z, Z^m, Z_m,
\tilde z)$. Now, define the following quantities:
\begin{equation}
\left[ \begin{array}{c} z \\ Z^m \\ Z_m \\ \tilde z \end{array} \right] := 
\tilde z\left[ \begin{array}{c} U \\ Y^m \\ Y_m \\ 1 \end{array} \right]
\end{equation}
After tedious though straightforward algebra, one finds that the system of
equations
\[
U   &=& N(Y^m) \,, \non[.5ex]
Y^m &=& (Y_m)^\# 
\]
is invariant under the induced \7. These equations define a 28-dimensional
curved submanifold \cM in $\mathbf{56}$. Furthermore, it can be shown
that $(Y^m, \tilde z)$ form a system of Poincar\' e coordinates on \cM.
The 27-dimensional boundary is recovered as the set $\{ \tilde z = \infty\}$,
thus making the parallel with the AdS story complete.

\begin{figure}\label{fig:j3o}\begin{center}
\leavevmode
%
%
\begin{picture}(0,0)%
\includegraphics{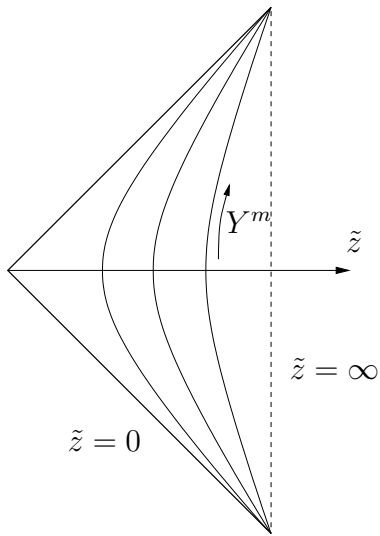}%
\end{picture}%
\setlength{\unitlength}{2072sp}%
\begingroup\makeatletter\ifx\SetFigFont\undefined
\def\x#1#2#3#4#5#6#7\relax{\def\x{#1#2#3#4#5#6}}%
\expandafter\x\fmtname xxxxxx\relax \def\y{splain}%
\ifx\x\y   
\gdef\SetFigFont#1#2#3{%
  \ifnum #1<17\tiny\else \ifnum #1<20\small\else
  \ifnum #1<24\normalsize\else \ifnum #1<29\large\else
  \ifnum #1<34\Large\else \ifnum #1<41\LARGE\else
     \huge\fi\fi\fi\fi\fi\fi
  \csname #3\endcsname}%
\else
\gdef\SetFigFont#1#2#3{\begingroup
  \count@#1\relax \ifnum 25<\count@\count@25\fi
  \def\x{\endgroup\@setsize\SetFigFont{#2pt}}%
  \expandafter\x
    \csname \romannumeral\the\count@ pt\expandafter\endcsname
    \csname @\romannumeral\the\count@ pt\endcsname
  \csname #3\endcsname}%
\fi
\fi\endgroup
\begin{picture}(4129,6344)(1329,-6833)
\put(3961,-3211){\makebox(0,0)[lb]{\smash{\SetFigFont{12}{14.4}{rm}{\color[rgb]{0,0,0}$Y^m$}%
}}}
\put(5401,-3436){\makebox(0,0)[lb]{\smash{\SetFigFont{12}{14.4}{rm}{\color[rgb]{0,0,0}$\tilde{z}$}%
}}}
\put(4726,-4966){\makebox(0,0)[lb]{\smash{\SetFigFont{12}{14.4}{rm}{\color[rgb]{0,0,0}$\tilde{z}=\infty$}%
}}}
\put(2071,-5821){\makebox(0,0)[lb]{\smash{\SetFigFont{12}{14.4}{rm}{\color[rgb]{0,0,0}$\tilde{z}=0$}%
}}}
\end{picture}
%
%
\caption{Embedding of $\cM$ in the \rep{56} representation of \7}
\end{center} \end{figure}

\mathversion{bold}
\section{Conformal duality symmetries}\label{sec:sugras}
\mathversion{normal}

The maximal supergravity theories in dimensions $2\le d\le 10$ can be obtained
by toroidal dimensional reduction of the $N\!=\!1$
supergravity~\cite{CrJuSc78} in eleven dimensions. After dualisation of some
of the $n$-form fields they all obey a global $E_{11-d}$ symmetry. This
symmetry was first discovered by Cremmer and Julia in the four-dimensional
theory~\cite{CreJul79} and was named ``hidden symmetry''. An exhaustive
treatment of all dimensions $3\le d\le 10$ can be found in Ref.~\cite{CJLP98}.
In the following, we will apply our construction to the five- and
four-dimensional maximal supergravity theories. However, we believe that similar
constructions exist in all other dimensions.

\subsection{Global symmetries of $d\!=\!5$, $N\!=\!8$ supergravity}
\label{sec:d5sugra}

$N\!=\!8$ supergravity in 5 dimensions~\cite{Crem80} is described by the
fields $e_\mu{}^\alpha$, $\psi_\mu^a$, $A_{\mu}^{ab}$, $\chi^{abc}$,
$\phi^{abcd}$ which transform under the \USp $R$-symmetry and are
antisymmetric and traceless in their indices $a,b,\ldots=1,\ldots,8$.

It is well-known that this theory admits a global \6 symmetry which leaves the
Lagrangian invariant. \6 acts on the vector fields $A_{\mu}^{m}
(m=1,\ldots,27)$ in the \rep{27} representation and on the scalars which can
be grouped together into a 27-bein $\cV_{m}{}^{ab}\in\6/\USp$ in the
\rep{\overline{27}} representation. The fields are summarised in
Table~\ref{d5fields}.

\begin{table}\label{d5fields}
\[
\begin{array}{llccc}
&&\text{spin}&\begin{array}{c}\USp\\\text{(local)}\end{array}
             &\begin{array}{c}\6\\\text{(global)}\end{array}\\[2ex]
\text{the graviton:}         &e_\mu{}^\alpha  &2  &\rep{1} &\rep{1}\\[.5ex]
8\,\text{gravitini:}        &\psi_\mu^a      &3/2&\rep{8} &\rep{1}\\[.5ex]
27\,\text{vector fields:}    &A_{\mu}^{m}     &1  &\rep{1} &\rep{27}\\[.5ex]
48\,\text{spin-1/2 fermions:}&\chi^{abc}      &1/2&\rep{48}&\rep{1}\\[.5ex]
42\,\text{scalars:}          &\cV_{m}{}^{ab}  &0  &\rep{27}&\rep{\overline{27}}
\end{array}
\]
\caption{Field content of $d\!=\!5,N\!=\!8$ supergravity}
\end{table}
The bosonic part of the Lagrangian is given by
\[
\cL &=& 
 -\ft{1}{4} e\,R 
 +\ft{1}{24}e\,\partial_\mu G_{mn} \partial^\mu (G^{-1})^{mn}
 -\ft{1}{8} e\,g^{\mu\rho}g^{\nu\sigma} G_{mn} F^m_{\mu\nu}F^n_{\rho\sigma}
\non[.5ex]
&&
 +\ft{1}{12}\epsilon^{\mu\nu\rho\sigma\lambda}F^m_{\mu\nu}F^n_{\rho\sigma}
  A^p_{\lambda} C_{mnp}\,,
\]
where $G_{mn}$ is the metric derived from the vielbein $\cV_{m}{}^{ab}$ and
$C_{mnp}$ are the coefficients of the norm $N_3$ (see Appendix~\ref{app:e6}).

In this \6-covariant formulation, the hidden global \6 symmetry is manifest.
There is an analogous form of the Lagrangian in all dimensions $3\le d\le
10$~\cite{CJLP98}. However, it is only in odd dimensions $d$ that the global
symmetry group $E_{11-d}$ be effectively realised on the Lagrangian; in even
dimensions the hidden symmetry is only a full symmetry of the equations of
motion.

Besides this global \6, an additional scaling symmetry $D$ was discovered to
exist in all maximal supergravities. This so-called ``trombone
symmetry''~\cite{CLPS98} finds its origin in the scaling symmetry of the
eleven-dimensional supergravity theory.  In the dimensionally-reduced
theories, the vielbein scales linearly and all $n$-index potential forms have
scaling weight $n$; scalar fields are left invariant:
\[
  e_\mu{}^\alpha \longrightarrow \lambda\,  e_\mu{}^\alpha\,, \quad
  A_{\mu}        \longrightarrow \lambda\,A_{\mu\nu\rho}\,,\;
  (\mu,\alpha = 0,\ldots,4)\ . \label{5d-trombone}
\]
Under $D$, the Lagrangian scales as $\cL\rightarrow\lambda^3\cL$.

In Ref.~\cite{CLPS98}, the authors pointed out the following problem: in the
five-dimensional {\em quantum} theory, the symmetries are expected to be
broken to a discrete subgroup by the Dirac quantisation condition. As to $\6
\times D$, one could, in principle, restrict both factor {\em separately} 
to discrete
subgroups. Since the only discrete subgroup in the second factor is $\{\pm
1\}$, the scaling would not survive quantisation.

In view of this problem, it looks desirable to find a larger group 
${\cal G} \supset \6 \times D$ such that both factors are {\em no longer
independent} but rather, are united in a nontrivial way dictated by the group
structure of $\cal G$: upon restriction to $\cal G(\mathbb{Z})$, the initially
independent quantisations of \6 and $D$ would be naturally related. 
A glance at \Eq{e7-grading} suggests an immediate (minimal) candidate 
for $\cal G$, namely \7. This is one piece of motivation for our quest for \7
in $d=5$ maximal supergravity. 

What are the minimal requirements that such an \7 realisation must meet?
\begin{enumerate}\label{conditions}
\item To begin with, $\6 \times D$ should be embedded in such a way that the
standard known action is retrieved upon restriction from \7. 
\item Secondly, upon dimensional reduction, the action must be such that it
reduces to the standard {\em linear} \7 action on $\mathbf{56}$ upon
dimensional reduction to $d=4$. If met, this condition would guarantee a
natural identification of the \7 in five-dimensions with the four-dimensional
(familiar) \7. In that case, the duality symmetry would be lifted, indeed,
up by one dimension.  
\end{enumerate}

\mathversion{bold}
\subsection{Extending the global symmetries beyond \6}\label{sec:e7realisation}
\mathversion{normal}
At this stage, we introduce an auxiliary 27-dimensional space $\cM$, with
coordinates $Y^m$. Let us temporarily {\em assume} that the fields listed in
Table~\ref{d5fields} depend on $Y^m$. In other words, the fields now live in 
the auxiliary 27-dimensional space $\cM$. Bold a step as this may seem at first sight, it
will be motivated by verifying that the minimal requirements of the
previous section can effectively be fulfilled. For now, this hypothesis
allows us to derive the required transformation properties of the supergravity
fields in $d=5$. From the last column in Table~\ref{d5fields}, the scalar
coset 27-bein $\cV_{m}{}^{ab}$ may be viewed as a one-form on $\cM$, while the
27 vector fields $A_{\mu}^{m}$ transform as the components of a vector field
on $\cM$. The remaining supergravity fields are scalars on $\cM$.

From tensor calculus, the following transformation rules for arbitrary tensor
fields $T(Y)^{m}$ and $T(Y)_{m}$ are induced by the \6 action on $\cM$, as
given in \Eq{e7-conf}:
\[
\delta_{78} T^{m}(Y) &=& {\Lambda^n}_p Y^p \partial_n T^m 
                        +{\Lambda^m}_n T^n\,,\non[.5ex]
\delta_{78}\, T_{m}(Y) &=& {\Lambda^n}_p Y^p \partial_n T_m 
                          +{\Lambda_m}^n T_n\,; \label{e6-field-trafo}
\]
the first terms on the r.h.s.\ are the transport terms and the
parameters of \6 are related as $\Lambda_m{}^n=-\Lambda^m{}_n$. For mixed-type
and higher-order tensors, one derives similar transformation rules.

The second terms in the r.h.s. of \Eq{e6-field-trafo} characterise $T^m$ and
$T_m$ as \rep{27} and \rep{\overline{27}} {\em linear} representations of \6.
This is the actual behaviour that we wish to recover. For this to occur, the
transport terms in \Eq{e6-field-trafo} must vanish. Therefore, in order to
satisfy the first condition on p.~\pageref{conditions}, we conclude that the
fields can at most depend on $Y$ via the \6-invariant norm $N_3(Y)$:
\[ 
e_\mu{}^\alpha (Y) &=& e_\mu{}^\alpha\,(N_3(Y))\,, \non[.5ex]
A_\mu^m (Y)        &=& A_\mu^{m}\,(N_3(Y))\,,\non[.5ex]
\cV_{m}{}^{ab}(Y)  &=& \cV_{m}{}^{ab}\,(N_3(Y))\,.
\]

How about the scaling behaviour under $D$? As tensors, $T(Y)^{m}$ and $T(Y)_{m}$
behave as follows under scale transformations: 
\[
\delta_{1} T^{m}(Y) &=& H\,Y^n\partial_n T^m + H\, T^m\,,\non[.5ex] 
\delta_{1}\,  T_{m}(Y) &=& H\,Y^n\partial_n T_m - H\, T_m\, .
\]
Matching this with the scaling laws in \Eq{5d-trombone}, we are to 
conclude that, initially,
\[ 
e_\mu{}^\alpha (Y) &=& N_3(Y)\, e_\mu^0{}^\alpha \non[.5ex] 
A_\mu^m (Y)        &=& N_3(Y) A_\mu^{0\,m} \label{Y-dependence}
\]
where the fields $e_\mu^0{}^\alpha$, $A_\mu^{0\,m}$, and also the scalars
$\cV_m{}^{ab}$ do not depend on $Y$. The resulting Lagrangian is
\[\label{Y-lagrangian} 
\cL(Y) = N(Y)^3 \cL_0 
\]
where $\cL_0$ is independent of $Y$ and only the spacetime-global factor in 
front of $\cL_0$ will change under reparametrisations in the $Y$-space. 
In particular, the group \7 which is realised as a conformal group on the 27
coordinates $Y^m$ only affects the global prefactor $N(Y)^3$ and thus 
becomes a genuine symmetry at the level of the equations of motion.

Finally, the conformal action of \7 on $\cM$ given in \Eq{e7-conf}, 
is found to induce the transformation rules below on tensors:
\[
\delta_{\overline{27}} T^{m}(Y) &=& 
                         \ft12 F_n K^{pn}{}_{qr} Y^q Y^r \partial_p T^m 
                        + F_n K_p{}^n{}_q{}^m Y^q T^p\ ,\label{field-trafo-1} \\
\delta_{\overline{27}}\, T_{m}(Y) &=& 
                           \ft12 F_n K^{pn}{}_{qr} Y^q Y^r \partial_p T_m 
                        + F_n {K^{pn}{}_{qm}} Y^q T_p\,. \label{field-trafo-2}
\]
with parameters related by $K_p{}^n{}_q{}^m=-K^{pn}{}_{qm}$.
 
\subsection{Some remarks on conformal dualities}
\label{sec:remarks}
The proposed realisation of \7 deserves some further comments. 

First, the form of the Lagrangian, \Eq{Y-lagrangian}, is a product of two
factors: a constant $\cL_0$ (that is, as far as the $Y$-dependence is
concerned), and a prefactor $N^3(Y)$. Under the action of Diff$(\cM)$,
this {\em Lagrangian} would generically transform into $f[N(Y)] \cL_0$;
since this may be viewed as a global rescaling of the initial Lagrangian, one
might be tempted to infer that the equations of motion remain invariant under
arbitrary reparametrisations Diff$(\cM)$.  However, there is a catch: the
factorised form of $\cL$ resulted from the particular $Y$-dependence of all
tensor fields via $N(Y)$ solely. For
\begin{equation}
N(Y)^3 \cL_0 \rightarrow f[N(Y)] \cL_0
\end{equation} 
to be induced from the transformations of individual {\em tensor fields}, one
must be careful not to destroy the specific dependence of the fields via
$N(Y)$.  General diffeomorphisms on the space $\cM$ would generically introduce
an explicit field dependence on all 27 $Y^m$-coordinates; as such, 
factoring out $N(Y)$
(like in \Eq{Y-dependence}) would no longer be possible. The only admissible
coordinate transformations on $\cM$ therefore, are those which leave the norm
$N(Y)$ conformally invariant. This singles out \7, embedded in Diff(\cM) as 
proposed in \Eq{e7-conf}.

Next, it is noteworthy that our proposed realisation on $\cV_{m}{}^{ab}(Y),
A_{\mu}^{m}(Y)$ is {\em linear} in the fields. This feature has two important
implications:
\begin{enumerate}
\item the space-time gauge invariance of the vector fields is manifestly
preserved; nonlinear transformations not spoiling gauge invariance are not
evident.
\item upon dimensional reduction to $d=4$, the vector fields in $\rep{27}$
combine with the Kaluza--Klein vector from the space-time metric (and their
duals) into the linear $\rep{56}$ of \7; had our construction been nonlinear in
the fields, demonstrating that things do work out consistently would be far
harder a task. 
\end{enumerate}

Thirdly, inspection of \Eq{field-trafo-1} and \Eq{field-trafo-2} reveals that the
space on which the symmetry is realised is ${\cal W} = \Phi \otimes
\mathbb{C}[Y^m]$, $\Phi$ being the space of ($Y$-independent) supergravity
fields and their space-time derivatives; further, $\mathbb{C}[Y^m]$ is the
polynomial ring in 27 variables. The latter has the structure of a graded
space,
\begin{equation}
\mathbb{C}[Y^m] \simeq \bigoplus_{n=0}^{\infty} {\rm Sym}^n(V)\ ,
\end{equation}
where $\cM$ is a 27-dimensional abstract vector space. In other words, the 
complete polynomial ring is graded by the degree of homogeneous polynomials.
Accordingly, we decompose ${\cal W} = \oplus_{n} {\cal W}_n$ in an obvious
notation. The nice feature of this fact is that the transformation rules for
$Y$-dependent fields show that the grading is respected by the realisation,
i.e.,
\begin{equation}
  0
  \begin{array}{c}\mbox{\raisebox{-1.2ex}[0pt][0pt]{$\longleftarrow$}}\\
                  \mbox{\raisebox{0ex}[0pt][0pt]{
  $\rep{\scriptstyle{27}}$}}\end{array} 
  \Phi 
  \begin{array}{c}\mbox{\raisebox{-2ex}[0pt][0pt]{
  $\rep{\scriptstyle{\overline{27}}}$}}\\
                  \mbox{\raisebox{-1ex}[0pt][0pt]{$\longrightarrow$}}\\
                  \mbox{\raisebox{1ex}[0pt][0pt]{$\longleftarrow$}}\\
                  \mbox{\raisebox{2ex}[0pt][0pt]{
  $\rep{\scriptstyle{27}}$}}\end{array} 
  {\cal W}_1
  \begin{array}{c}\mbox{\raisebox{-2ex}[0pt][0pt]{
  $\rep{\scriptstyle{\overline{27}}}$}}\\
                  \mbox{\raisebox{-1ex}[0pt][0pt]{$\longrightarrow$}}\\
                  \mbox{\raisebox{1ex}[0pt][0pt]{$\longleftarrow$}}\\
                  \mbox{\raisebox{2ex}[0pt][0pt]{
  $\rep{\scriptstyle{27}}$}}\end{array} 
  {\cal W}_2
  \begin{array}{c}\mbox{\raisebox{-2ex}[0pt][0pt]{
  $\rep{\scriptstyle{\overline{27}}}$}}\\
                  \mbox{\raisebox{-1ex}[0pt][0pt]{$\longrightarrow$}}\\
                  \mbox{\raisebox{1ex}[0pt][0pt]{$\longleftarrow$}}\\
                  \mbox{\raisebox{2ex}[0pt][0pt]{
  $\rep{\scriptstyle{27}}$}}\end{array} 
  \cdots \ .
\end{equation}
In particular, the dilatation $(\mathbf{1})$ and \6 $(\mathbf{78})$ do not
shift degrees, hence map $\Phi$ into itself, while the $\mathbf{27}$
translations annihilate the degree-0 space. $\Phi$ implements $\mathbf{27} =
\mathfrak{g}^{-1}$ trivially.  Put equivalently, only the $\mathfrak{g}^0
\oplus \mathfrak{g}^{+1}$ subalgebra of \7 is seen to act nontrivially on the
supergravity fields.

Finally, let us comment on the physical (in)significance of introducing
$Y$-dependence in the supergravity fields. As explained in section
\ref{sec:e7realisation}, the particular way in which quantities are made 
$Y$-dependent effects in an initial $N(Y)^3$ prefactor in front of the
lagrangian. Acting with \7 on $N(Y)^3 {\cal L}_0$ could rescale $N(Y)$, at
worst. If so, however, the scale factors could be absorbed into the coupling
constant in front of the {\em action}:
\begin{equation}
S = \frac 1 {{\ell}_p^3} \int e {\cal L}
\end{equation}
(where $\ell_p$ is the Planck length in five dimensions). Namely, $\ell_p
\rightarrow \ell_p(N(Y))$ would result in a $Y$-dependent coupling, and the
conformal duality transformations, \Eq{field-trafo-1}, would map a theory with
a given $\ell_p$ to one with a possibly different ${\ell'_p}'$. That is, after
fixing an arbitrary choice of a point $Y$. The picture suggested is thus:
there is a one-dimensional space ${\cal T}$ of $N\!=\!8$ theories,
distinguished by different values of $\ell_p$. Rather than by $\ell_p$, one
may choose to parametrise this line by $N(Y)$; the 27-dimensional space of
$Y^m$ is thus viewed as a highly redundant description of $\cal T$: the
one-parameter family of 26-dimensional subspaces $\{ N(Y) = N_0\}$ in the
(carrier space of) $\mathbf{27}$ reflect this redundancy.

\mathversion{bold}
\subsection{Duality symmetries of $d\!=\!4$, $N\!=\!8$ supergravity}
\label{sec:d4sugra}
\mathversion{normal}

We want to discuss how to extend the global \7 symmetry in four dimensions to
\8. For this purpose we need a realisation of \8 on a $56$-dimensional vector
space.

The problem in this case is the fact that \8 does not admit a
three-grading\Eq{3-grading}. However, it admits a 5-graded decomposition
w.r.t.\ its subgroup $E_{7(7)} \times \SLR[2]$. Denoting its Lie algebra by
$\mathfrak{e}_8$ we have
\[
\begin{array}{c@{\,\,\oplus\,\,}c@{\,\,\oplus\,\,}c@{\,\,\oplus\,\,}c@{\,\,\oplus\,\,}c}
\mathfrak{g}^{-2} & \mathfrak{g}^{-1} & \mathfrak{g}^0 
                  & \mathfrak{g}^{+1} &\mathfrak{g}^{+2}  \\[1ex]
\rep{1}  & \rep{56} & (\rep{133}\oplus\rep{1}) & \rep{56} & \rep{1} 
\end{array}
\label{5-grading}
\]
An important property of this decomposition is the fact that the subspaces of
grade $-1$ and $-2$ together form a maximal Heisenberg subalgebra. The
corresponding generators $X^{ij},X_{ij}\in\mathfrak{g}^{-1}$ and
$x\in\mathfrak{g}^{-2}$ obey the commutation relations
\[\label{heis-sub}
 {}[X^{ij},X_{kl}] &=& -2 \,\delta^{ij}_{kl} x \,.
\]

In Ref.~\cite{GuKoNi00} a realisation of \8 on the $(56+1)$-dimensional
Heisenberg subalgebra $\mathfrak{g}^{-1} \oplus \mathfrak{g}^{-2}$ has been
constructed. This so-called quasiconformal realisation is similar to the
conformal realisation of \7 in the sense that an \7 invariant norm
$N_4(X^{ij},X_{ij},x)$ on that $57$-dimensional space is conformally invariant
under the \8 action. The norm is given by
\[
 N_4(X^{ij},X_{ij},x) = \cI_4(X^{ij},X_{ij}) - x^2
\]
where $\cI_4$ is the quartic invariant of \7 in the \rep{56} representation.
Enforcing the \8-invariant condition $N_4=0$, or equivalently
\[
 \cI_4(X^{ij},X_{ij}) = x^2\,,
\]
eliminates the 57th variable $x$ and yields a realisation of \8 on $\Rn^{56}$.
Following the same procedure as in section~\ref{sec:e7realisation}, we can
realise \8 on the fields of maximal supergravity in $d\!=\!4$ and the
Lagrangian will by construction again only change by a global factor.

\section{Conclusions}\label{sec:conclusions}

In this paper, we have shown how to lift the continuous hidden-symmetry group
\7 from $d=4$ to $d=5$. In five dimensions, the symmetries are conformally
realised in the sense of section~\ref{sec:remarks}, where the idea of
``trombone symmetry'' was borrowed from Ref.\cite{CLPS98}: rather than $\6
\times D$ there, in our approach the entire \7 emerges.

The presented construction hinges largely on the fact that \7 may be viewed as
a conformal extension of \6. As explained in section~\ref{sec:remarks}, a
``natural'' way to lift the \7, whilst preserving the known \6, is to make the
supergravity fields live, but not propagate, in a 27-dimensional space $\cal
M$, on which \7 acts by (generalised) conformal transformations.  In the
context of hidden symmetries, this auxiliary space $\cal M$ had already
appeared in the literature in Ref.~\cite{WitNic00}, although in a different
guise.

A geometrical picture of the situation illustrated how \7, acting in the
$\mathbf{56}$ linear representation, ends up realised as nonlinear coordinate
transformations on $\cM \hookrightarrow \mathbf{56}$. Rather than \7,
the continuous group of hidden symmetries, it is believed that only the U-duality discrete subgroup
$\7(\mathbb{Z}) \subset \7$ survives quantisation \cite{HulTow95}. In
principle, using this embedding, the U-dualities of $d\!=\!4, N\!=\!8$
supergravity can now be lifted to the five dimensional theory.

In summary, the results presented in this paper may be said to give additional
insight in the origin of four-dimensional symmetries from a five-dimensional
perspective. As to higher dimensions, we believe that our construction can be
generalised with minor modifications to account for the different structure of
the U-duality groups. 

\ifthenelse{\boolean{ack}}{
\bigskip
\noindent
{\bf\large Acknowledgements}\medskip

\noindent This work was supported in part by the EU network `Superstrings'
(HPRN-CT-2000-00122) and by the PPARC special grant `String Theory and
Realistic Field Theory' (PPA/G/S/1998/0061). The authors are also supported by
a PPARC fellowship. The authors are grateful to B.~Brinne and S.~Silva for
helpful discussions. One of the authors (K.K.) would also like to thank
M.~G\"unaydin, H.~Nicolai, and P.~West for enlightening discussions on related
topics.
}{} 

\pagebreak
\begin{appendix}
\mathversion{bold}
\section{Properties of \6}\label{app:e6}
\mathversion{normal}

The Lie algebra of \6 has maximal compact subalgebra \USp.  We denote tensors
transforming in the fundamental \rep{27} transfomation by $T^m$
$(m=1,\ldots,27)$ and tensors transforming in \rep{\overline{27}} by $T_m$. An
\6 transformation in the \rep{27} representation is given by
\[
\delta T^m = \Lambda^m{}_n T^n \quad\text{and}\quad
\delta T_m = \Lambda_m{}^n T_n
\]
with a traceless matrix $\Lambda_m{}^n=-\Lambda^m{}_n$. 

The tensor product
\[ 
\rep{27} \times \rep{\overline{27}} = \rep{1} + \rep{78} + \rep{650}
\]
contains a singlet $\delta^m_n$, whereas 
\[ \label{27x27}
\rep{27} \times \rep{27} &=& 
  \rep{\overline{27}} + \rep{351} + \rep{351^\prime}\non[.5ex]
\rep{\overline{27}} \times \rep{\overline{27}} &=& 
  \rep{27} + \rep{351} + \rep{351^\prime}
\]
does not. This is reflected by the fact that \6 does not possess a quadratic
invariant. Howewer, there is an invariant $N_3$ of third order:
\[
 N_3 (T) = T^m T^n T^p C_{mnp}
\]

The \rep{27} can be associated with the 27-dimensional exceptional Jordan
algebra $J_3^{\On_S}$ where $\On_S$ denotes the split real form of the
octonions $\On$. This Jordan algebra possesses a symmetric Jordan product $X
\circ Y$ with invariance group \4 and a triple product\footnote{The
  corresponding formula for the triple product in~\cite{GuKoNi00} should read
  correctly $\{X,Y,Z\}^{ab} = 16 (X\oversym{{}^{ac}Z_{cd}Y^{db}}
  +Z\oversym{{}^{ac}X_{cd}Y^{db}}) +4 X^{ab}Y^{cd}Z_{cd} +4 Y^{ab}X^{cd}Z_{cd}
  +4 Z^{ab}X^{cd}Y_{cd}+4 \Omega^{ab}(X^{cd}Y_{de}Z^{ef}\Omega_{cf})\,$.}
\[
\{\;,\;,\;\}\,: \rep{27}\times\rep{\overline{27}}\times\rep{27}
 \longrightarrow \rep{27}
\]
which is invariant under \6. This triple product is crucial in the
construction of \7 as the conformal extension of \6. We denote its structure
constants by $K^{mn}{}_{pq}$:
\[
 \{X,Y,Z\}^m = K^{mn}{}_{pq}\, X^p Y_n Z^q
\]

Using\Eq{27x27} one can also define a ``conjugations'' ${}^\#$ which maps
\rep{27} into \rep{\overline{27}} and vice versa. It is quadratic in the sense
that $(\lambda T)^\#=\lambda^2 T^\#$ and obeys the relation
$T^{\#\#}=N(T)\,T$.

\ifthenelse{\boolean{internal}}{
\section{Translation from the old formulas:}

The norm:
\[ 
N(X) &=& X^{ab}\Omega_{bc}X^{cd}\Omega_{de}X^{ef}\Omega_{fa} \non
     &=& - X^a{}_b X^b{}_c X^c{}_a 
\]

The conjugation:
\[ 
(X^{ab})^\# &=& -\ft12\, X^{ac}\Omega_{cd}X^{db} 
                +\ft1{16} \Omega^{ab} (X^{cd}X_{cd})
\]
}{} 

\end{appendix}
%
%
\providecommand{\href}[2]{#2}\begingroup\raggedright\endgroup
%
%
\end{document}